\documentclass[prx,twocolumn,aps,showpacs,nofootinbib]{revtex4} 

\usepackage{amsmath}
\usepackage{graphicx}
\usepackage{dcolumn}
\usepackage{bm}
\usepackage{color}
\usepackage[normalem]{ulem}

\newcommand{\ep}{\epsilon}

\usepackage{color}

\begin{document}

\title{The three faces of entropy for complex systems -- information, thermodynamics and the maxent principle}

\author{Stefan Thurner$^{1,2,3,4}$, Bernat Corominas-Murtra$^{1}$, and Rudolf Hanel$^{1,4}$
}

\affiliation{
$^1$ Section for the Science of Complex Systems, CeMSIIS, Medical University of Vienna, 
Spitalgasse 23, A-1090, Vienna, Austria\\
$^2$ Santa Fe Institute, 1399 Hyde Park Road, Santa Fe, NM 87501, USA\\
$^3$ IIASA, Schlossplatz 1, 2361 Laxenburg, Austria \\
$^4$ Complexity Science Hub Vienna, Josefst{\"a}dterstrasse 39, A-1090 Vienna, Austria \\
}

\date{Version \today}

\begin{abstract}
There are three ways to conceptualize {\em entropy}: 
entropy as an extensive thermodynamic quantity of physical systems (Clausius, Boltzmann, Gibbs), 
entropy as a measure for information production of ergodic sources (Shannon), and  
entropy as a means for statistical inference on multinomial Bernoulli processes (Jaynes maximum entropy principle). 
Even though these notions are fundamentally different concepts, the functional form of the entropy 
for thermodynamic systems in equilibrium, 
for ergodic sources in information theory, and 
for independent sampling processes in statistical systems, is  degenerate, $H(p)=-\sum_i p_i\log p_i$. 
For many complex systems, which are typically history-dependent, non-ergodic and non-multinomial,
this is no longer the case. Here we show that for such processes 
the three entropy concepts lead to different functional forms of entropy.  
We explicitly compute these entropy functionals for three concrete examples. 
For {\em P{\'o}lya urn processes}, which are simple self-reinforcing processes,  
the source information rate is $S_{\rm IT}=\frac{1}{1-c}\frac1N \log N$, 
the thermodynamical (extensive) entropy is $(c,d)$-entropy, $S_{\rm EXT}=S_{(c,0)}$, 
and the entropy in the maxent principle (MEP) is $S_{\rm MEP}(p)=-\sum_i \log p_i$.
For {\em sample space reducing} (SSR) processes, which are simple path-dependent processes 
that are associated with power law statistics, 
the information rate is $S_{\rm IT}=1+ \frac12 \log W$,  
the extensive entropy is $S_{\rm EXT}=H(p)$, and 
the maxent result is $S_{\rm MEP}(p)=H(p/p_1)+H(1-p/p_1)$. 
Finally, for {\em multinomial mixture} processes, the information rate is given by the conditional entropy $\langle H\rangle_f$, with respect to the mixing kernel $f$,
the extensive entropy is given by $H$, and the MEP functional corresponds one-to-one to the logarithm of the mixing kernel.
\end{abstract}

\pacs{
05.20.-y, 
05.40.-a, 
89.75.Da,
05.40.Fb,
02.50.Ey
}

\maketitle

\section{Introduction}

Historically, the notion of entropy emerged in conceptually distinct contexts. 
In physics thermodynamic entropy $S$ has been introduced by Clausius as an 
extensive quantity that links temperature with heat \cite{carnot,clausius}. 
Boltzmann could relate this thermodynamic entropy to the number of microstates $W$ in a system, 
\begin{equation}
 	S_{\rm B}=k_{\rm B}\ln W \quad ,  
\end{equation}
assuming that in equilibrium all microstates are equally probable \cite{kittel}. 
Microstates are different configurations that e.g. ideal gas particles or spins can assume 
as samples of a Bernoulli process in a closed volume at constant internal energy. 
In the case that microstates appear with probabilities $p_i$, the  entropy functional reads
\begin{equation}
	H(p)=-k_{\rm B}\sum_{i=1}^W p_i\log p_i \quad , 
	\label{shannon}
\end{equation}
which is often referred to as the Gibbs formula. We set $k_{\rm B}=1$ in the following.
Obviously, Eq. (\ref{shannon}) is an {\em extensive} functional in the number of states $W$. 
Since Boltzmann we identify the extensive functional with thermodynamic entropy,  $S_{\rm EXT}=H(p)$.
This notion gave way to the success story of statistical mechanics. 

Independently from physics, in the context of {\em information theory} (IT), 
a functionally identical notion of entropy appears  \cite{Shannon1948,Kraft1949,McMillan1956}. 
There $H$ quantifies how efficiently a particular stream of information can be coded, 
if the information source  is an ergodic finite state machine with $W$ states. 
The information production rate $S_{\rm IT}=H$ determines if information can be coded, 
transmitted through a noisy channel and decoded in an error-free way. 

In the attempt to formulate statistical physics in a way that is independent of the physics of particles or spins, the 
{\em maximum entropy principle} (MEP) was developed \cite{Jaynes1957}. It is a way to address statistical 
inference problems that are not confined to physics. Again the same functional $H$ appears, $S_{\rm MEP}=H$.
The MEP approach is explicitly grounded in the statistics of multinomial Bernoulli processes.
It can be used to infer the so-called {\em maximum configuration} from  particular data, i.e. the distribution $p_i$ of states $i$ that is 
the most likely to be observed and that dominates the overall behavior of a system. 

What these three very different approaches have in common is that 
physical processes behind ideal gases, information production of ergodic sources, and multinomial statistics, 
are all essentially Bernoulli processes. 
As we will show below, the particular functional form of $H$ from Eq. (\ref{shannon}) is a generic consequence 
of this fact. For this reason the different entropy concepts appear degenerate in the sense that $S_{\rm EXT}$, $S_{\rm IT}$, and $S_{\rm MEP}$ 
all are expressed by the identical functional $H$. 
Processes that are non-ergodic, history-dependent, or that have long-term memory, 
explicitly break this degeneracy, which demonstrates that $H(p)$ is by no means 
a universal functional that fits all purposes. 
We will show in detail that the three concepts behind entropy lead to distinct entropy 
functionals that have to be determined for every family of processes individually.

We introduce some notation. For physical systems one typically uses 
the configuration-, for IT, the process picture. They are equivalent. We will use both. 
By $X$ we denote a class of systems or of processes. 
A configuration in a physical system corresponds to a path that a process can take;  
paths are the microstates in the process picture.
A class is parametrized by a set of parameters $\theta$, we write $X(\theta)$. 
For example, the class $X$ of Bernoulli processes is given by the prior probabilities $q_i$, and 
$\theta=(q_1,\cdots,q_W)$. {\em Sample space} is denoted by $\Omega=\{1,2,\cdots,W\}$.  
$W=W(X(\theta))$ is the number of distinct elements in $\Omega$.
Sequences  $x(N)=(x_1,\cdots,x_N)\in\Omega^N$ are either 
{\em paths} sampled by a process of length $N$, $X(N,\theta)=(X_1,X_2,\cdots,X_N)$ 
or configurations of a system with $N$ elements, $X(N,\theta)$ with $W(X(N,\theta))=W^N$. 
We can distinguish $W^N$ different paths that a process $X(N)$ can take, 
or $W^N$ distinct configurations of a system $X(N)$\footnote{
	For Bernoulli processes this equivalence is trivial. It is equivalent to toss $N$
	independent dice at once, or to toss one die $N$ times in a sequence. 
	Both result in $N$ i.i.d. random variables $X_n$, 
	$n\in{1,\cdots,N}$. The process- and the system picture only differ in terms of 
	how variables $X_n$ may depend on other variables $X_m$. For processes there may exist a time ordering,
	where $X_n$ depends on variables $X_m$ that appeared earlier in time, $m<n$.}. 
The {\em histogram} of a sequence $k(x(N))=(k_1,\cdots,k_W)$  
keeps track of how often state $i$ is visited in the sequence $x(N)$. $p_i=k_i/N$ 
are the relative frequencies and $p=(p_1,\cdots,p_W)$ is the distribution of relative frequencies.
The {\em phasespace volume} is the number of configurations that a system 
at a given resolution can be in.  

Section \ref{sec1} introduces the three entropy concepts in their respective contexts. 
Section \ref{sec2} shows that entropies are degenerate for Bernoulli processes, section \ref{sec3} deals with 
P\'olya urn processes and derives their corresponding IT, thermodynamic (extensive) and the MEP entropies. 
Sections \ref{sec4} and  \ref{sec5} do the same for sample-space reducing processes and 
multinomial mixture models, respectively. Section \ref{sec6}  concludes.

\subsection{The three concepts of entropy \label{sec1}}

In the following we discuss how the three notions of entropy arise in the contexts of 
information theory, thermodynamics, and the maximum entropy principle.

\subsubsection{The information theoretic entropy and entropy rate}

Shannon's approach to information theory deals with the question of how many 
bits per letter are needed on average to transmit messages of a certain type through 
information channels and what happens if these channels are noisy. 
Consider an information source processes $X(\theta)$. The sample space $\Omega$ in this context 
is called an {\em alphabet} of letters (or a lexicon of words) $i \in \{1,\cdots W\}$.  
$X(N,\theta)$ generates messages, i.e. realizations or samples $x(N)=(x_1,x_2,\cdots, x_N)$. 
For transmitting the message through an information channel one has to translate messages 
into a code that has $b$ symbols in the code-alphabet (typically code is binary, $b=2$) 
such that the average number of bits per letter, becomes minimal. 
Intuitively this means that frequently observed letters are assigned short binary 
codewords while infrequent letters are assigned longer codewords.

Shannon identified four properties of a functional $H$ -- the four Shannon-Khinchin (SK) axioms -- 
that measures the average amount of information (in bits) that is required to encode messages that generate 
letters $i$ with probabilities $p_i$. 
Three of these properties are of technical nature, 
SK axiom 1: $H$ is a continuous function that depends on $p$ only and no other variables; 
SK axiom 2: $H(p_1,\cdots,p_W)$ is maximal for the uniform distribution $p_i=1/W$; 
SK axiom 3: $H(p_1,\cdots,p_W,0)=H(p_1,\cdots,p_W)$. 
The fourth property, the so-called {\em composition axiom} (SK axiom 4) states, 
that $H$ measures information independent of the way the $W$ states get sampled with the probabilities $p_i$. 
It states that a system composed of two systems $A$ and $B$ that are statistically dependent  the 
entropy of the composed system $S(AB) = S(A) + S(B | A)$ is the entropy of system $A$ alone plus the  
entropy  of system $B$, conditional on $A$.  Details on conditional entropy follow below. 
SK axioms 1-4 determine $H$ uniquely up to a multiplicative constant. $H$ is the functional given in Eq. (\ref{shannon}). 

Two theorems, one by Kraft \cite{Kraft1949} one by and McMillan \cite{McMillan1956} assure that there 
exists  a practical family of uniquely decodable codes (the prefix codes) if and only if $\sum_{i\in\Omega}b^{-\ell_i}\leq 1$,
where $\ell_i$ is the length of the codeword for letter $i$ and $b$ is the size of the code alphabet. 
For a binary code $b=2$ this means that if the source variables $X_n(\theta)$ are identically independently distributed (i.i.d.), or equivalently, 
if letters $i$ appear with fixed probabilities $p_i$ for all $n$,  
one can find codewords of length $\ell$ such that, $1-\log_2(p_i)\geq \ell_i\geq -\log_2(p_i)$. 
Such a code requires the least number of bits for transmitting messages. Using $\log_2(p_i)=\log(p_i)/\log 2$ we have,  
\begin{equation}
	1+H(p)/\log 2\geq \langle\ell\rangle \geq H(p)/\log 2\quad . 
\end{equation}
This means that $H(p)$ establishes the lower bound for the 
so-called {\em information rate} or source information rate 
of  i.i.d. processes in bits per letter for prefix codes. 

What if we are not encoding letters but entire parts of messages $x(N)$ that are sampled from $\Omega^N$ 
with respective probabilities $p(x(N))$? The information rate of $x(N)$ is generally defined as \cite{Thomas:2001}, 
\begin{equation}
	S_{\rm IT}(x(N))=-\frac1N\log p(x_N,x_{N-1},\cdots,x_2,x_1)\quad ,  
\label{entropyrate0}
\end{equation}
where the joint distribution appears. 
For processes, where each $X_n$ may depend on earlier events, we can re-write Eq. (\ref{entropyrate0}).
Using the notions for the empty sequence $x(0)=\emptyset$ and for the initial distribution $p(i|\emptyset)$,
we write $p(x(N))=\prod_{n=1}^Np(x_n|x(n-1))$, and obtain, 
\begin{equation}
	S_{\rm IT}(x(N))=-\sum_{n=1}^N \log p(x_n|x(n-1))\quad .
\label{entprod1}
\end{equation}
The Shannon-McMillan-Breiman (SMB) theorem \cite{Shannon1948,McMillan1953,Breiman1957} 
states that for Markov chains with transition probabilities $p(i|j)$ and stationary distributions $p_j$
the asymptotic information rate is given by the {\em conditional entropy} $H(X_{n+1}|X_{n})$, i.e. 
\begin{equation}
	\lim_{N\to\infty} S_{\rm IT}(x(N))=-\sum_{j=1}^W p_j \sum_{i=1}^W p(i|j)\log p(i|j) \quad .
\label{equipartprop}
\end{equation}
For Bernoulli processes, where $p(i|j)=p_i$, obviously 
$\lim_{N\to\infty} S_{\rm IT}(x(N))=H(p)$.
Note that for history-dependent process classes $X$, the law of large numbers that plays a crucial role in the SMB theorem, 
does not necessarily apply and the situation needs to be analyzed carefully for each specific path-dependent process. 

The SMB theorem states that for Markov chains one can transmit messages at lower
bit rates, $H(X_{n+1}|X_{n})\leq H(p)$, by using optimal code lengths $\ell_i(j)\sim -\log_2 p(i|j)$ that are
conditioned on the most recent event $j$ of a message, $\langle\ell\rangle\sim H(X_{n+1}|X_{n})/\log 2$. 
Also history-dependent processes can in principle be coded more efficiently. 
However, this does not mean that the transmission of information becomes more efficient since the 
key (decoding table) to the constantly up-dated coding schemes must be transmitted in addition to the 
source information. The {\em effective information rate} measures the total amount of information the 
sender has to transmit to the receiver.

\subsubsection{Thermodynamics and extensive entropy}

Traditionally thermodynamics deals with ``homogeneous'' matter, such as ideal gases or solid 
bodies in thermal equilibrium and characterizes systems independent of size, shape and scale
in terms of so-called {\em intensive} variables, such as temperature and pressure. 
Conjugate variables, such as volume and entropy, relate the intensive variables 
to the number of system components, or more precisely, to the number of the degrees of freedom. 
If extensive variables do not scale linearly with the degrees of freedom, no reasonable thermodynamic equations 
will exist. 

If two initially separated systems  $A$ and $B$ -- that are at the same temperature and pressure, 
with volumes $V_A$ and $V_B$ and thermodynamic entropies $S(A)$ and $S(B)$, respectively -- 
are combined, this implies that $V_{AB}=V_A+V_B$, and  $S(AB) =  S(A) + S(B)$.
The extensivity of the thermodynamic entropy results from particles being {\em indistinguishable}, 
meaning that permutations of indistinguishable particles do not change the microstate. This effectively resolves the 
Gibbs paradox by constraining particles to their independent share of the volume $V/N$, see for example \cite{vanKampen}. 

Assume that $W=\bar{W}(X_n)$ is the number of states the $n$-th particle can be in, say discrete positions in a container. 
Then, if $N_A$ and $N_B$ are the numbers of identical particles in the two containers, respectively,  
one finds that the {\em effective number of configurations} 
$\hat{W}$ in the combined container is given by $\hat{W}(AB)=W^{N_A+N_B}=W^{N_A}W^{N_A}=\hat W(A)\hat W(B)$. 
Boltzmann entropy $S_{\rm B}=\log \hat W=N\log W$ is extensive in $N$.
In the case that the states that each particle can be in are sampled from a given distribution $q$ 
-- which may not be uniform -- one can still estimate the {\em effective} number of states
as $\hat{W}\sim e^{NH(q)}$, where $H(q)$ is the Gibbs formula Eq. (\ref{shannon}) for distribution $q$. 
As a consequence, $e^{H(q)}$ measures the effective amount of states per particle\footnote{ 
	Alternatively, one can measure the first moment of the rank $r(i|q)$ 
	of states $i$ with respect to the distribution function $q$. 
	The rank $r(i|q)$ is a permutation on $\Omega$, such that $r(i|q)>r(j|q)$ if $q_i>q_j$.
	For a reference process with being concentrated uniformly on $\bar W$ states
	one finds $\langle r\rangle_n\equiv\sum_{i=1}^W q_i(n)r(i|q(n))=(\bar W+1)/2$. 
	Conversely, one may define $\bar W(X_n)\equiv 2\langle r\rangle_n-1$. 
}, and Boltzmann entropy remains extensive, $\log \hat{W}=NH(q)$. 

This is generally valid for systems or processes $X(N)=(X_1,\cdots,X_N)$ described by i.i.d. variables $X_t$. 
Systems or processes with strong constraints, strong interactions, with non-stationary prior
probabilities for states $q_i(t)$, strong internal correlations or with history-dependent dynamics, 
typically populate subspaces of the entire phasespace and $H$ (Gibbs formula)  is no-longer extensive. 
For examples see e.g. \cite{HTphasespacegrowth,HTentropy,HTMGM3}. 
In the more general case, one can estimate $\hat W(N)$ by, 
\begin{equation}
	\hat W(N)\equiv \prod_{t=1}^N \bar W(X_t) \quad ,
	\label{WRq}
\end{equation}
where, again, we measure $\bar W(X_t)\sim e^{H(q(t))}$. 
Such systems or processes are called non-extensive and the SK axiom 4 (composition axiom) is violated. 
In this case $H$ lost the extensive property. However, one can find a  functional expression for 
an entropy that remains extensive -- even though the underlying system or process is non extensive. 
We call such a functional the {\em extensive entropy}, $S_{\rm EXT}$.
Since from equation (\ref{WRq}) it follows that $\hat W(N)$ is monotonically increasing in $N$, 
an inverse function $L_X$ exists such that $L_X(\hat W(N))=N$, and a unique extensive trace-form 
functional can be found, see Appendix \ref{apA}, 
\begin{equation}
	S_{\rm EXT}(p)=\sum_{x\in\Omega^N}s(p(x)) = Ns_0 \quad . 
	\label{traceform}
\end{equation}
Here $p(x)$ is the probability to sample path $x$ and $s_0$ is a constant. 

For classes $X$, that are compatible with the first three Shannon-Khinchin axioms SK axiom 1-3, but violate
SK axiom 4, -- often non-ergodic processes -- all extensive entropies $S_{\rm EXT}$ can be classified 
by $(c,d)$-entropies,  \cite{HTclassification}. 
These, in a convenient representation, take the form
\begin{equation}
	S_{(c,d)}(p) = \frac{\frac{e}{c}\sum_{i=1}^W\Gamma(1+d,1-c\log(p_i))-1}{1-c+cd}\quad .
\label{scd}
\end{equation}
$S_{(c,d)}$ is parametrized by two scaling exponents $c$ and $d$ that characterize the asymptotic scaling 
behavior of the entropy of the non-extensive system or process.
The exponents are one-to-one related with the phasespace of the system \cite{HTphasespacegrowth},
and can be computed using
\begin{equation}
	\begin{array}{lcl}
	\frac1{1-c}&=&\lim_{N\to\infty} N\frac{d}{dN}\log \hat W(N)\\
	d&=&\lim_{N\to\infty} \log \hat W(N)\left(c-1+\frac{1}{N\frac{d}{dN}\log \hat W(N)}\right)\quad .
	\end{array}
	\label{ass2}
\end{equation}
$(c,d)$-entropies are extensive quantities for non-extensive system classes.

Extensive systems correspond to the special case $c=1$ and $d=1$,  and one finds 
$\frac1e S_{(1,1)}(ep)=-\sum_i p_i \log p_i$ ($e$ is the Euler constant). 
The special case of $d=0$ corresponds to power laws and recovers Tsallis entropy \cite{tsallisbook},  
$\frac1\eta S_{(c,0)}(\eta p)=(1-\sum_i p_i^c)/(c-1)$, where $\eta=c^{1/(c-1)}$ (note that $\lim_{c\to 1}\eta=e$). 
For $c<1$, $(c,d)$-entropies describe the phasespace growth of so-called winner-takes-all processes (WTA), 
where probabilities $p_i$ of sampling states  $i\in\Omega$ concentrate over time in one single element $j\in\Omega$, 
the winner, and  $\lim_{n\to\infty} \bar W(X_n)=1$. WTA processes also violate SK axiom 3.

\subsubsection{The entropy of the maximum entropy principle}

The {\em maximum entropy principle} (MEP) is tightly related with the question 
of finding the most likely observable macroscopic property (macrostate) of a system or a process. 
The distribution function $p$, or the histogram $k$, of events $x_n$ that occurred along the 
path $x$ of a process $X(N,\theta)$, is such a macrostate. 
In other words, how do we find the most likely distribution function of a given process or a system?
Denoting the probability of finding the histogram by $P(k|\theta)$, 
the most likely histogram $k^*$ is obtained by maximizing $P(k|\theta)$ with respect to $k$ under the constraint, $\sum_{i} k_i=N$.
$k^*$ is the best predictor for observing a macrostate that is generated by the process $X(N,\theta)$.  
If $P$ becomes sharply peaked as $N$ becomes large, predictions will become very accurate. 

The MEP of the process $X(N,\theta)$ is obtained by factorizing $P$ into two terms, $P(k|\theta) =M(k)G(k|\theta)$.
$M$ is the {\em multiplicity} of the macrostate $k$, the number of microstates that lead to the macrostate.
$G(k|\theta)$ is the probability of a microstate belonging to $k$. 
Whenever such a factorization can be defined in a meaningful way, a corresponding maximum entropy principle exists.

Taking logarithms $\log P = \log M + \log G$ does not change the location $k^*=k$ of the maximum of $P(k|\theta)$, 
and, 
\begin{equation}
	\underbrace{ \frac 1f \log P(k|\theta)}_{-S_{\rm rel}}= 
	\underbrace{ \frac 1f \log M(k) }_{S_{\rm MEP}} + 
	\underbrace{ \frac 1f \log G(k|\theta) }_{-S_{\rm cross}}\quad .
	\label{MEP1}
\end{equation}
Here $f$ is an appropriate scaling factor, which corresponds to the degrees of freedom of microstates, see \cite{HTMGM3}.

$S_{\rm rel}$ is the {\em relative entropy} or {\em information divergence}. 
Note that for Bernoulli processes, where $\theta$ is given by the prior probabilities $q$, $S_{\rm rel}$ is identical to the Kullback-Leibler divergence \cite{KullbackLeibler},  
$H_{\rm rel}(p|q)\equiv D_{\rm KL}(p||q)=\sum_i p_i (\log p_i-\log q_i)$.

$S_{\rm MEP}= \frac 1f \log M(k)$ is the entropy that appears in the MEP, we call it {\em MEP-entropy}. 
It is sometimes called the {\em reduced} Boltzmann entropy\footnote{Boltzmann's 
	principle as formulated by Planck \cite{Planck1901}   
	identifies entropy $S_{\rm B}$ with the logarithm of multiplicity, $S_{\rm B} =  k_{\rm B} \log M$.
}. 
which is defined as $s_B=S_B/f$. 

$S_{\rm cross}(p|\theta)= -\frac 1f \log G(k|\theta)$ is the {\em cross-entropy},
which depends on sets of constraints imposed by the parametrization $\theta$. 
Again, for Bernoulli processes with prior probabilities $q$, the cross entropy takes the well known form,   
\begin{equation}
	H_{\rm cross}(p|q)=-\sum_{i=1}^W p_i\log q_i\quad .
\label{crossent}
\end{equation}

\section{Bernoulli processes \label{sec2}}

We compute the three entropies, $S_{\rm IT}$, $S_{\rm EXT}$, and $S_{\rm MEP}$ 
for Bernoulli processes and show that they are identical with $H$ from Eq. (\ref{shannon}). 
Bernoulli processes have no memory and states $i=1,\cdots,W$
are sampled independently from the prior probability distribution $q=(q_1,\cdots,q_W)$.
Bernoulli processes of length $N$, $X(N,\theta)$ are parametrized by $\theta\equiv q$. 

Consider the histograms $k$ with $\sum_{i=1}^W k_i=N$, as the macrostates of the Bernoulli process
and sequences $x(N)$ as their microstates, then the probability to sample a particular sequence $x(N)$ 
with histogram $k$ is given by $G(k|q)=\prod_{i=1}^W q_i^{k_i}$. The multiplicity $M(k)$ 
is given by the multinomial factor $M(k)={N\choose k}$, and the probability to sample histogram $k$ is $P(k|q)=M(k)G(k|q)$.
The number of degrees of freedom of a sequence of length $N$ is $f=N$.

\subsection{The information rate of Bernoulli processes}

Since Bernoulli processes have no memory the transition probabilities $p(i|x,\theta)=q_i$ do not depend on path $x$.
The information rate from Eq. (\ref{entprod1}) is,
\begin{equation}
	\begin{array}{lcl}
		S_{\rm IT}(x)&=&-\frac1N\sum_{n=1}^N\log p(x_n|\,x(n-1),\theta) \\
		&=&-\frac1N\sum_{i=1}^Wk_i\log q_i\\
		&=&-\sum_{i=1}^W p_i\log q_i
		=H_{\rm cross}(p|q)\quad .
	\end{array}
\end{equation}
Since $\lim_{N \to \infty} p=q$, for a typical sequence $x$ one finds,
\begin{equation}
		\lim_{N\to\infty} S_{\rm IT}(x)=H_{\rm cross}(p|q)=H(p) \quad . 
\end{equation}
The entropy $S_{\rm IT}=H(p)$ measures the typical information rate for optimally coded Bernoulli processes.

\subsection{The extensive entropy of Bernoulli processes}

There are three ways to see what the extensive entropy of Bernoulli process $X(N,\theta)$ is. 

(1) Since Bernoulli processes fulfil all 4 Shannon Khinchin axioms, a well known theorem by Shannon \cite{Shannon1948} (appendix II) states that  
$S_{\rm EXT}(p)=H(p)$. 

(2) The effective number of configurations $\hat W(N)=\bar W(X(N))$ of a Bernoulli process $X(N)$ 
grows exponentially, $\hat W(N)=\bar W^N$. This is because $X(N)$ is composed of $N$ i.i.d. Bernoulli trials $X_n$. 
Using Eq. (\ref{extent}) and setting $\hat W(N)=\bar W^N$,  we see that $L_X(y)=\log y /\log \bar W$. 
As a consequence $s_0=\log \bar W$, and $s(y)=-y\log y$, meaning that $S_{\rm EXT}(p)= \sum_i s(p_i)= H(p)$.

(3) Using Eq. (\ref{ass2}) and the exponential  phasespace growth, $\hat W(N)=\bar W^N$ of Bernoulli processes 
one verifies that $c=1$ and $d=1$. For obtaining $c$ one computes,  
\begin{equation}
	\frac1{1-c}=\lim_{N\to\infty} N\frac{d}{dN} N\log \bar W = \infty \quad .
\end{equation}
As a consequence $c=1$. Similarly one obtains $d=1$. Since $S_{(1,1)}=H$, we conclude that $S_{\rm EXT}=H$.

\subsection{The MEP-entropy of Bernoulli processes \label{BernMEP}}

Since for Bernoulli processes the degrees of freedom are simply given by the number of samples $f=N$, 
using Stirling's approximation $N!\sim N^Ne^{-N}$ it is easy to see,  
\begin{equation}
	\begin{array}{lcl}
		S_{\rm MEP}&=&\frac1N\log{N\choose{k}}\\
		({\rm Stirling})&\sim&\frac1N\log\frac{N^N}{\prod_{i=1}^W k_i^{k_i}} =-\frac1N\log\prod_{i=1}^W p_i^{k_i}\\
		&=&-\sum_{i=1}^W p_i\log p_i =H(p)\quad.
	\end{array}
\end{equation}
The maxent entropy $S_{\rm MEP}$ of Bernoulli processes is again equivalent to $H(p)$. 

The relative entropy $S_{\rm rel}=-\frac1f\log P$ is given by the Kullback-Leibler divergence $D_{\rm KL}$,  
\begin{equation}
	S_{\rm rel}(p|\theta)=\sum_{i=1}^W p_i(\log p_i -\log q_i)\equiv D_{\rm KL}(p||q)\quad .
\end{equation}

The  cross entropy $S_{\rm cross}=-\frac1N\log G$ is given by $-\sum_i p_i\log q_i$ and 
imposes a linear first moment constraint on $p$ in the MEP. 
This can be seen by re-parametrizing $q_i$ by $\exp(-\alpha-\beta\ep_i)$, which yields  
$S_{\rm rel} \sim H(p)-\alpha\sum_{i=1}^Wp_i-\beta\sum_{i=1}^W p_i\ep_i$. 
$\alpha$ and $\beta$ play the role of Lagrangian multipliers in the maximization problem. 
For Bernoulli processes the maximum configuration asymptotically predicts $p^*=q$.

\section{The three entropies of P{\'o}lya urn processes \label{sec3}}

\subsection{P{\'o}lya urn processes}

\begin{figure}[t]
\centering
	\includegraphics[width=0.8\columnwidth]{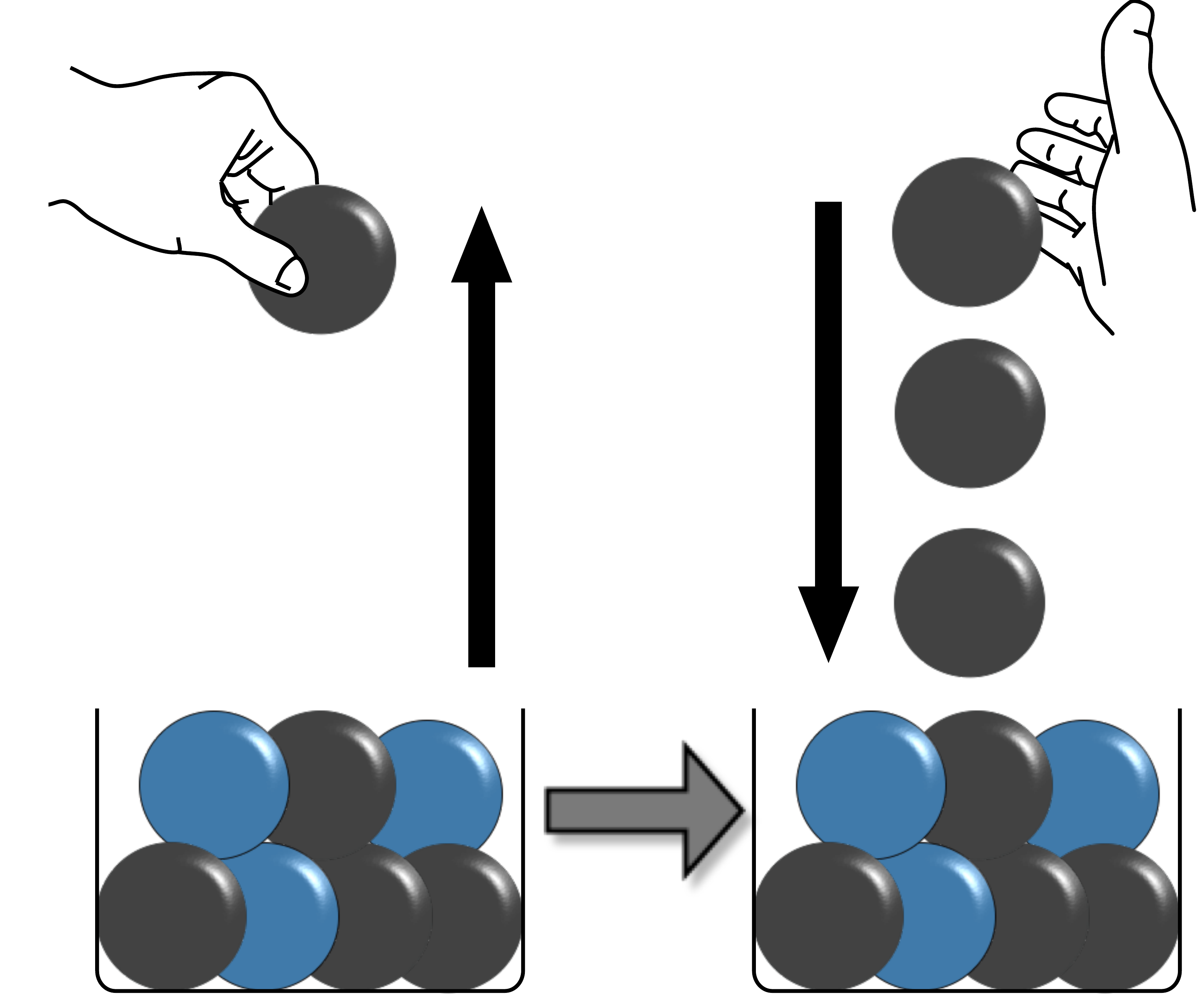}
\caption{  
Schematic illustration of a P{\'o}lya urn process. When a ball of a certain color is drawn, 
it is then replaced by $1+\delta$ balls of the same color (here $\delta =2$). The process 
is repeated $N$ times. This reinforcement process creates a history-dependent dynamics. 
After \cite{HCMTPolya2016}.
\label{fig:polya}
}
\end{figure}

Multi-state P{\'o}lya urn processes \cite{Polya1923,Polya1930} are an abstract representation 
of path-dependent, self-reinforcing processes with memory.
A P\'olya urn is initially filled with $a_i$ balls of color $i=1,\cdots,W$. One draws the first ball of color 
$x_1=i$ with probability $q_i=a_i/A$, where $A=\sum_{i=1}^W a_i$ is the total number of balls initially in the urn. 
If we draw a ball of color $i$, we do not only replace it, as we would do in a Bernoulli process (drawing with replacement), 
but we add another $\delta$ balls of the same color and thus reinforce the probability to draw color $i$ in 
subsequent trials, see Fig. \ref{fig:polya}. As a consequence, the probability to draw another 
state (color) $i$ after $N$ samples drawn is given by, 
\begin{equation}
	p(i|k,\theta)=\frac{a_i + k_i\delta}{A+N\delta}=\frac{q_i + k_i\gamma}{1+N\gamma}\quad ,
\label{polyatransp}	
\end{equation}
where $\gamma=\delta/A$, is the {\em reinforcement} parameter, $q_i=a_i/A$, and
$\theta=(q,\gamma)$ is the set of parameters characterizing the process. 
If $\gamma=0$ the P{\'o}lya urn process is just `drawing with replacement'  
and the same as a Bernoulli process. If $\gamma>0$, 
the probability to draw color $i$ in the $N+1$ th sample depends 
on the history of samples $x$ in terms of the histograms $k$. 

P{\'o}lya urn processes and non-linear versions of it exhibit a cross-over between dynamics 
that is asymptotically a Bernoulli processes (weak reinforcement)
and a dynamics that is referred to as ``winner takes all'' (WTA) dynamics (strong reinforcement). 
For intermediate reinforcement strengths $\gamma$, whether the system behaves 
one way or the other  depends on the details of random events that
happened early on in the process. How sequences $x=(x_1,\cdots,x_N)$
behave for large $N$ depends on the samples $x_n$, that are drawn at times 
$n$ much smaller than $N$.

P\'olya processes operate at the edge of Bernoulli- and WTA dynamics.
If we measure the histogram $k(N_1)$ of the process after $N_1$ 
steps, we may continue the process by thinking of starting a different P\'olya 
urn with an initial condition $k(N_1)$. For this we consider the histogram $k'=k-k(N_1)$ of 
$N'=N-N_1$ samples and define $a_i(N_1)=a_i + k_i(N_1)\delta$ and $A(N_1)=A+\delta N_1$. 
It is easy to see that one again is looking at a P{\'o}lya urn process. 
However, the parameters have been modified from 
$\theta=(q,\gamma)$ to $\theta'=(q',\gamma')$, where,   
\begin{equation}
	\begin{array}{lcl}
		q'&=&\frac{q_i + \gamma k_i(N_1)}{1+\gamma N_1}\gamma\\
		\gamma'&=&\frac{\gamma}{1 + N_1\gamma}\quad .
	\end{array}
	\label{laterq}	
\end{equation}
As a consequence, the effective reinforcement $\gamma'<\gamma$, and $\gamma'(N_1)\to 0$ as $N_1\to\infty$. 
The distribution $q'$ gets modified by the  history of the process $x(N_1)$ and the effective reinforcement 
parameter $\gamma'$ decreases over time. Whether or not a particular realization of a process defined by 
$\theta$ enters the WTA dynamics therefore depends on whether the modified 
P{\'o}lya urn with parameters $\theta'$, enters WTA dynamics or not. This depends on
which path $x(N_1)$ the urn process took within the first $N_1$ steps. If in those first steps  
one of the elements $i$ acquires most of the weight, the process can enter WTA dynamics,
meaning that  $i$ eventually gets sampled almost all of the time. 
Non-linear P\'olya processes, where the effective reinforcement decays more slowly as time progresses, 
almost certainly enter WTA dynamics.  
We can now discuss the three entropies of P{\'o}lya processes.

\subsection{The information rate of P{\'o}lya urn processes}

In WTA scenarios the relative frequency $p_j$ to observe the winner $j$ approaches $1$, 
meaning that $p$ concentrates on the winner state $j$, which essentially becomes the only state that is sampled,  
\begin{equation}
	p_j(N)\sim 1-\frac{1-q_j}{1+\gamma N}\quad.
	\label{winner}
\end{equation}
Without knowing the exact distribution of the ``looser'' states $i\neq j$
we assume that all those states have equal probabilities, 
\begin{equation}
	p_i(N)=\frac{1-p_j(N)}{W-1}=\frac{1-q_j}{(W-1)(1+\gamma N)}\quad.
	\label{looser}
\end{equation}
Following  Eq. (\ref{winner}) the information rate of a WTA process can be estimated,
\begin{equation}
	\begin{array}{lcl}
	NS_{\rm IT}(x)&=&-\sum_{n=1}^N \log p(x_n|x(n-1))\\
	&\sim&\frac{1-q_i}{\gamma}\log N + {\rm const}.\quad .
	\end{array}
\end{equation}
The information rate of a P{\'o}lya process in the WTA mode asymptotically approaches 
zero. The total information production, i.e. the number of
bits required to encode the entire sequence, grows logarithmically, $NS_{\rm IT}(x(N))\propto \log N$. 
If the P\'olya processes does not enter WTA dynamics it behaves like a Bernoulli process sampling from the limit distribution 
$p(\infty)=\lim_{N\to\infty} p(N)$ with a information rate, $H(p(\infty))$.

\subsection{The extensive entropy of P{\'o}lya urn processes}

The effective number $\bar W$ of a typical sequence $x$ of length $N$, with $j$ the winner in the WTA process, can be 
estimated by inserting inserting $q$ from Eq. (\ref{winner}) and Eq. (\ref{looser})
into $\bar W(n)\sim \exp(H(q(n)))$, or alternatively by using this $q$ to compute first rank moment 
$\langle r\rangle$ and $\bar W(n)\sim 2\langle r\rangle-1$. One uses Eq. (\ref{WRq}) to compute, 
\begin{equation}
	\hat W_{\mbox{P{\'o}lya}}(N) \propto \left(1+\gamma N\right)^{\frac{1-q_j}{\gamma}} \quad .
\label{polyaeffw}
\end{equation}
From Eq. (\ref{ass2}) it follows that $1/(1-c)=(1-q_j)/\gamma$. Therefore, the $(c,d)$-class of P{\'o}lya urn processes is 
\begin{equation}
	c=1-\frac{\gamma}{1-q_j}\quad{\rm and}\quad d=0\quad .
\end{equation}
Note that $c$ is negative for $\gamma$ sufficiently large, which means that the  SK axiom 3 is violated by P\'olya processes in WTA dynamics.
As a consequence (\ref{ass2}) might no-longer hold, since it was derived under the assumption that SK axioms 1-3 do hold. 
However, one can still safely compute the extensive entropy using Eq. (\ref{extent}) with and Eq. (\ref{polyaeffw}) to find,
\begin{equation}
	S_{\rm EXT}(q)=\frac{1}{1-q_j}\sum_{i=1}^W q_i\log_c q_i  = S_{(c,0)} \quad ,
\end{equation}
where $\log_c(x)=(x^{1-c}-1)/(1-c)$ and $c=1-\gamma/(1-q_j)$. This is exactly the result that we get from Eq. (\ref{ass2}).

\subsection{The MEP  of P{\'o}lya urn processes}

For P{\'o}lya urn processes the probability to observe a sequence $x$
is the same as the probability to observe any other sequences $x'$ with the same histogram $k$. 
Therefore $P(k|\theta)=M(k)G(k|\theta)$ factorizes into the multiplicity 
$M(k)$, which is given by the multinomial factor, and the sequence probability, $G(k|\theta)$. 
One might conclude that the number of degrees of freedom scales like $f=N$.
In this case the P{\'o}lya MEP is $H$ plus cross-entropy terms. 
If $\gamma$ is sufficiently small, this is indeed true and the P{\'o}lya processes essentially behave as Bernoulli processes. 
If $\gamma$ is sufficiently large however, the P{\'o}lya process is likely to enter the WTA dynamics if
one state gets sampled repeatedly in the very beginning of the process. 
How often on average do we expect state $i$ to be sampled in a row at the beginning of a P{\'o}lya process?
The answer is, 
\begin{equation}
	\langle n\rangle(q_i)=\sum_{n=0}^\infty n\left(1-\frac{q_i+\gamma n}{1+\gamma n}\right)
	\prod_{m=0}^{n-1}\frac{q_i+\gamma n}{1+\gamma n}\quad .
\end{equation}
To first order in $\gamma$ one can estimate that,  
\begin{equation}
	\langle n\rangle(q_i)\sim \frac{q_i}{1-q_i-\kappa(q_i)\gamma}\quad ,
\end{equation}
where $\kappa(q_i)>1-q_i$. 
As $\gamma\to (1-q_i)/\kappa(q_i)$ from below, $\langle n\rangle(q_i)\to\infty$.
This means that if states $j$ violate the condition 
$\gamma<(1-q_j)/\kappa(q_j)\leq 1$, it 
becomes likely that the P\'olya process enters the WTA dynamics. 
Practically this means that usually WTA behavior 
can be observed if a state $i$ gets sampled repeatedly within 
the first few steps of the P{\'o}lya process. Otherwise 
the effective reinforcement $\gamma'$ becomes too small  to  
enter the WTA dynamics and the sampling distribution $q(N)$ approaches that of a Bernoulli process. 

For sufficiently large $\gamma$, one finds the situation that 
$G(k|\theta)$ can be written as $G(k|\theta)=\tilde M(k)\tilde G(k|\theta)/M(k)$, 
so that $MG=\tilde M\tilde G$, \cite{HCMTPolya2016}. 
This means that the probability for the histogram $P=\tilde M\tilde G$, no-longer 
depends on the multinomial factor $M$ at all. One observes that for $\gamma>0$ 
the expression $\log \tilde M$ scales very differently than multinomial multiplicities.
With $f=1$ the MEP entropy $S_{\rm MEP}\equiv \frac1f\log \tilde M$ becomes a 
well defined {\em generalized} relative entropy, and $S_{\rm cross}=-\frac1f\log \tilde G$ 
a {\em generalized} cross entropy functional. 
In \cite{HCMTPolya2016} we have shown in detail that,
\begin{equation}
	\begin{array}{lcl}
	S_{\rm MEP}(p|N)&\sim&-\sum_{i=1}^W \log(p_i+1/N) \\
	S_{\rm cross}(p|q,\gamma,N)&\sim&-\frac1{\gamma}\sum_{i=1}^Wq_i\log(p_i+1/N)\quad . 
	\end{array}
\end{equation}
The numerical values for the WTA dynamics (one winner and $W-1$ losers) are 
\begin{equation}
	\begin{array}{lcl}
	S_{\rm MEP}&\sim&(W-1)\log N+ {\rm const}.\\	
	S_{\rm cross}&\sim&\frac{1-q_j}{\gamma}\log N + {\rm const}.\\	
	\end{array}
	 \label{polyaent2}
\end{equation}

The generalized relative entropy $S_{\rm rel}$ can also be viewed as 
the information divergence of P{\'o}lya processes, 
\begin{equation}
	S_{\rm rel} = D_{\rm P\acute{o}lya}(p|\theta)=\sum_{i=1}^W\left(\frac{q_i}{\gamma}-1\right)\log(p_i+1/N)\quad .
\end{equation}
$D_{\rm P\acute{o}lya}$ is convex in $p_i$ only if $\gamma<q_i$. The processes becomes unstable
if the reinforcement parameter $\gamma$ is sufficiently large.
This intrinsic instability of self-reinforcing processes makes MEP predictions 
of the distribution function $p=(p_1,\dots,p_W)$ unreliable since large deviations from the maximum 
configuration $p^*$ remain probable, even for large $N$. 
In other words, no well defined typical sets of paths $x$ form with respect to the distribution of states
$i\in\Omega$.
However, quite remarkybly, ensembles of P{\'o}lya urns show stable frequency and rank distributions.
If we want to predict the relative frequencies of states ordered according to their rank, 
the largest frequency having rank $r=1$, the second largest frequency rank $r=2$, etc., then this rank distribution 
$\tilde p=(\tilde p_1,\cdots,\tilde p_W)$,  can still be predicted with high accuracy \cite{HCMTPolya2016}. 
P{\'o}lya urn paths produce typical sets with respect to the most likely observed rank distribution.

\subsection{Summary P{\'o}lya urn processes}

P\'olya urn processes either enter WTA dynamics or behave as a Bernoulli process. 
For WTA scenarios one finds that $S_{\rm IT} \sim \frac{1}{N} \log N$, the extensive entropy is 
$S_{\rm EXT} = S_{c,0}$, where $c<0$, and $S_{\rm MEP} \sim(W-1)\log N$, Table \ref{table2}. 
The corresponding numerical  values of the different entropies yield similar results,
P\'olya urns  
\begin{equation}
	NS_{\rm IT} 
	\sim S_{\rm cross} \sim \frac{1-q_j}{(W-1)\gamma}S_{\rm MEP}\quad.  	
	 \label{polyaent3}
\end{equation}
Again, $S_{\rm cross}$ is a measure of information production. However, instead of measuring 
the information rate, which becomes zero, it measures the {\em total} information production.
This matches the intuition that the most likely ``winner'' is a state that happens to be ``in lead'' at the
very beginning of the process. With some non-vanishing probability another state can take over the lead 
within the first few steps. However, if this happens it becomes very unlikely 
that the P\'olya urn process can still enter the WTA dynamics because of  decreasing effective 
reinforcement parameter $\gamma'$. The process then asymptotically approaches
a Bernoulli process, where the three entropies are degenerate, 
$S_{\rm IT}\sim S_{\rm EXT} \sim S_{\rm MEP}\sim H$.

\section{The three entropies of sample space reducing processes \label{sec4}}

\subsection{Sample space reducing processes}

Sample space reducing (SSR) processes are processes whose sample space reduces as they evolve over time. 
They provide a way to explain the origin and ubiquity of power laws 
in complex systems, and Zipf's law in particular \cite{BRSstaircase,BRSstaircase2}. 
SSR processes are typically irreversible, dissipative processes that are driven between sources and sinks. 
Complicated driven dissipative processes such as sandpile dynamics, \cite{OsloPile}, 
can often be decomposed into simpler SSR processes. 
Examples of sample space reducing processes include fragmentation processes, 
sentence formation \cite{BRSstaircase2}, diffusion and search processes on networks \cite{CMHT16} 
and cascading processes \cite{CMHT17}. 

SSR processes can be viewed as processes where the currently occupied 
state determines the sample space for the next.
If the system is in state $i$ it can sample states from a sample space $\Omega_i$. 
Often sample spaces are nested along the process, meaning that $\Omega_i\subset\Omega_j \Leftrightarrow i>j$. 
In such cases, as the process evolves, the sample space successively becomes smaller. 
Eventually a SSR process ends in a sink state, $i=1$ ($\Omega_1$ is the empty set).
The dynamics of such systems is irreversible and non-ergodic. 
To keep dynamics going, SSR processes have to be restarted, which can lead to a 
stationary, driven, and irreversible process that is effectively ergodic. 

\begin{figure}[t]
	\centering
		\includegraphics[width=0.8\columnwidth]{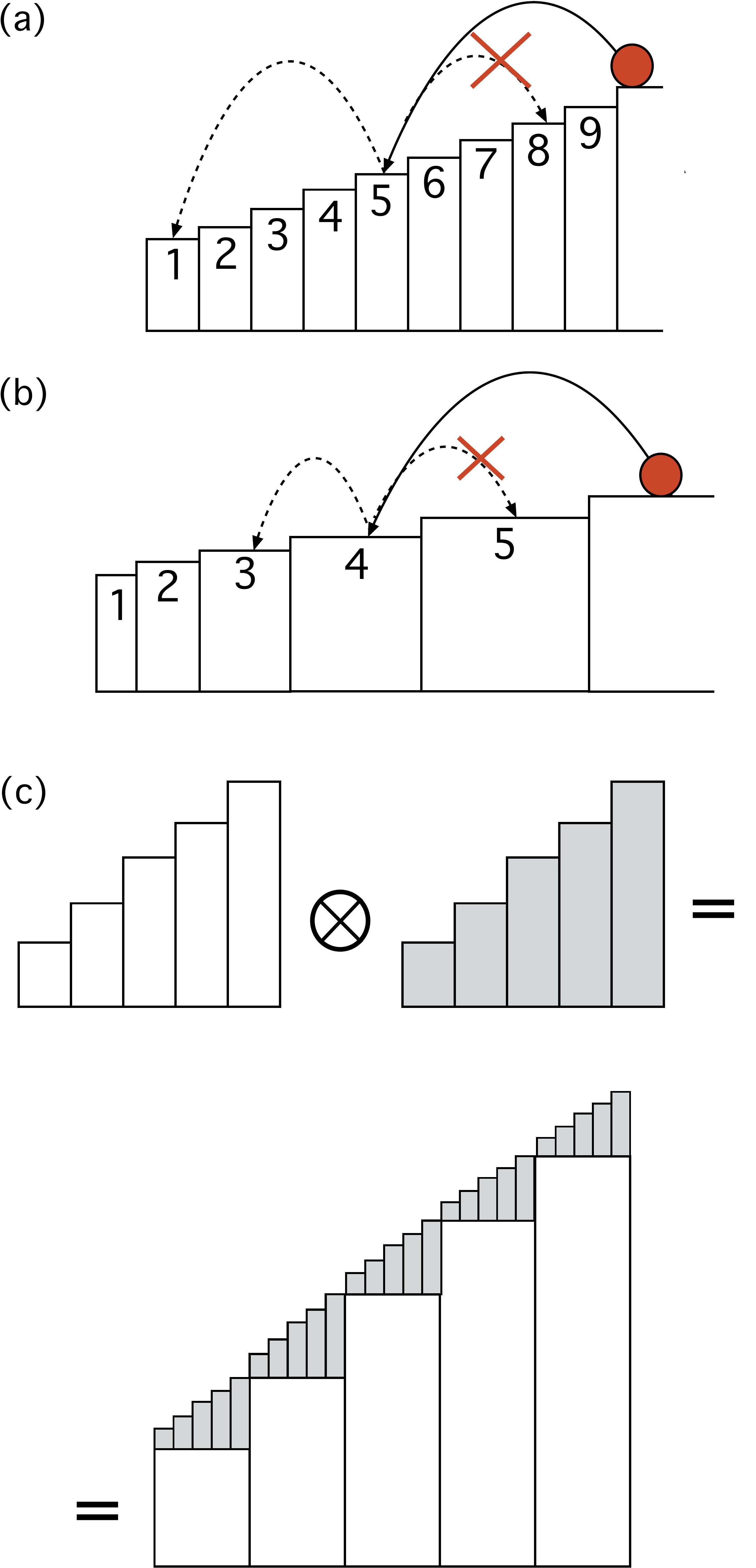}
	\caption{  
	(a) Pictorial view of a SSR process. A ball bounces downwards only, with random step sizes. After several
iterations of the process, the visiting probabilities of states $i$ approach $p_i = i^{-1}$ (ZipfÕs law). 
	(b) SSR with non-uniform prior probabilities. For a wide class of prior probabilities the visiting distributions still follow Zipf's law.
	(c) Combining two staircase processes through a  ``cartesian product''. 
	\label{fig:productstairs}
	}
\end{figure}

A simple way to depict a SSR process is a ball bouncing downwards random distances on a staircase. It never jumps upwards.  
Each stair represents a state $i$. State $i=1$ corresponds to the bottom, $i=W$ to the top of the staircase, 
see Fig. \ref{fig:productstairs} a. Obviously, successive sample spaces are nested. 
A ball on step $i$ can sample from all steps below itself $j<i$ with equal- or prior probabilities $q$.   
If the steps carry prior probabilities $q=(q_1,\cdots,q_W)$ 
(which can be intuitively interpreted as the widths of the steps, Fig. \ref{fig:productstairs} b) 
the process will visit state $j<i$ with probability
$q_j/Q_{i-1}$, where $Q_i=\sum_{s=1}^i q_s$ is the cumulative distribution of $q$ up to $i$.
Regardless of $q$ (exceptions are discussed in \cite{CMHT16}) the SSR processes 
still follows Zipf's law in the visiting distributions, $p_i = i^{-1}$. 
By restarting the process one forces the process to become quasi-ergodic, 
meaning that a stationary distribution $p$ exists, despite the process being 
irreversible. By allowing the process to jump to any position with a given frequency $1-\lambda$, the 
visiting distributions remain exact power laws, $p_i = i^{-\lambda}$, \cite{BRSstaircase,CMHT16}.
In the following we discuss the three entropies of the ``staircase process''.

\subsection{The information rate of SSR processes}

Note that SSR processes are Markov processes, the probability of sampling $x_n$ only depends on the previous sample $x_{n-1}$. 
Considering ensembles of ``staircases''  (restarting the SSR process every time it stops) 
allows us to treat the process as if it were ergodic and well defined asymptotic  
distributions $p=(p_1,\cdots,p_W)$ exist. The entropy production of typical sequences therefore yields,  
\begin{equation}
\begin{array}{lcl}
	S_{\rm IT}(x)&=&-\frac1{N}\sum_{n=1}^N \log p(x_{n}|x_{n-1})\\
&\sim&-\sum_{i,j=1}^W p(j|i)p_i\log p(j|i) =\sum_{i=1}^W p_i H_i\quad . 
\end{array}
\label{entprod2x}
\end{equation}
The entropy production of the SSR process is given by the {\em conditional entropy}. 
For uniform priors $q_i=1/W$ one computes the numerical value of the SSR entropy production, 
\begin{equation}
\begin{array}{lcl}
	S_{\rm IT}&=&p_1\log W+p_1\sum_{i=2}^W \frac{1}{i}\log i \\
	&\sim& 1+\frac12\log W + {\cal O}(1/\log(W))\quad .
\end{array}
\label{scprodX}
\end{equation}
Here we replaced sums $\sum_{i=a}^bf(i)$ by integrals $\int_{a-1/2}^{b+1/2}dx f(x)$.
Note that the 1 in Eq. (\ref{scprodX}) arises from the restarting procedure.

\subsection{The extensive entropy of SSR processes}

We quantify how the phasespace of a SSR process grows by the number of 
decisions (where the ball jumps next) the process takes along its path. A Bernoulli process on $W$ 
states chooses between $W$ possible successor states at every timestep. 
After $N$ samples the process selected one specific path among the $W^N$ possible. 

The effective number of decisions in a SSR process is computed using Eq. (\ref{WRq}).  
Note that by restarting the SSR process the process gets quasi-ergodic and that each state 
is visited with probability $p_i=p_1/i$, with $1/p_1=\sum_{i=1}^W 1/i$.
At state $i>1$ the process can sample from $W_i=i-1$ states
and restarting the process once it hits state $i=1$ means $W_1=W$ (it can jump anywhere).
With this we compute the typical size of phasespace, 
\begin{equation}
\begin{array}{lcl}
	\hat W(N)&\equiv&\prod_{n=1}^N W_{x_n}  \sim \prod_{i=1}^W W_i^{p_iN} = \bar W^N\quad .
	\label{WSSR}
\end{array}
\end{equation}
Consequently, $\bar W=W^{p_1}\prod_{i=2}^W (i-1)^{p_i}$, and
the average amount of choice per step involved in sampling a typical SSR sequence $x$ is
given by the numerical value
\begin{equation}
\log \bar W=\frac12\log W+ 1 + {\cal O}(1/\log W)\quad .
\end{equation}
The contribution of the constant $1$ comes from restarting the process. 
This implies that $\bar W \sim e\sqrt{W}>1$.

The definition of extensivity is tightly related with the way systems are composed. 
Staircase $A$ with $W(A)$ states can be combined with staircase $B$ with $W(B)$ steps, to a 
staircase $AB$ by substituting each step of staircase $A$
with a copy of staircase $B$, see Fig. \ref{fig:productstairs} c. We get,  
\begin{equation}
	\bar W(AB)=e\sqrt{W(A)W(B)}=\frac1e\bar W(A)\bar W(B)\quad .
\end{equation}
If we compose staircase $A$ $N$ times with itself we get $\bar W(A(N))=e(\bar W(A)/e)^N$. 
In other words, the quasi-ergodic SSR has an exponentially growing phasespace
and the extensive entropy is given by $S_{\rm EXT}=H$.

\subsection{The MEP of SSR processes}

To arrive at the MEP for SSR processes $X_{\rm SSR}$ with histogram $k=(k_1,\dots,k_W)$ as the macrostate, 
we need to determine the probability $P(k|q)=M(k)G(k|q)$ after $N$ observations of the process and 
determine the maximum configuration $k^*$ that maximizes $P(k|q)$. 
To compute $M$ we first decompose any sampled sequence 
$x=(x_1,\cdots,x_N)$ into shorter sequences $x^{r}$, such that $x=x^{1}x^{2}\cdots x^{R}$ is a 
concatenation of such shorter sequences. Any sequence $x^r$ is a sample of executing $X$ until $X$ stops. 
We refer to $x^r$ as one ``run'' of $X$. This means that any run $x^{r}=x^{r}_{1}x^{r}_{2}\cdots x^{r}_{N_r}$ 
is a monotonously decreasing sequence of states, $x^{r}_{n}>x^{r}_{n+1}$, ending in 
$x^{r}_{N_r}=1$, where $X$  stops and needs to be restarted. Note that $\sum_{r=1}^R N_r=N$. 
Since every run ends in state $1$ the number of runs equals the number of times state $1$ is sampled, $R=k_1$.
Arranging $x$ in a table with $W$ columns and $k_1$ rows, denoting a stair that gets 
visited by $*$ and a stair that does not get visited within a run by $-$, allows us to determine the probability 
$G$ and the multiplicity $M$ of a sequence $x$. 
\begin{equation}
	\begin{array}{c||c|c|c| c |c|c||}
	 r \times i  & W & W-1 & W-2 & \cdots & 2 & 1 \\
	\hline
	 1 & * & - & - & \cdots & * & * \\
	 2 & - & * & * & \cdots & - & * \\
	 3 & * & - & * & \cdots & - & * \\
	 \vdots & \vdots & \vdots & \vdots &  & \vdots & \vdots \\
	 R-2 & - & * & * & \cdots & - & * \\
	 R-1 & - & * & - & \cdots & * & * \\
	 R & - & - & * & \cdots & - & * \\
	\hline
	  & k_W & k_{W-1} & k_{W-2} & \cdots & k_2 & k_1 \\
	\end{array}
	\label{table1}
\end{equation}
We can directly assess the number $M$ of 
sequences $x$ that have the same histogram $k$. Note that 
column $i$ is $k_1=R$ entries long and contains $k_i$ items; $k_i\leq k_1$.
Therefore one can produce all those sequences $x$  
by re-arranging $k_i$ visits to state $i$ in a column of $k_1$ possible 
positions. Each column $i>1$ therefore contributes  to $M$ 
with the binomial factor ${k_1 \choose{k_i}}=k_1!/k_i!/(k_1-k_i)!$. 
As a consequence one finds $M(k)=\prod_{i=2}^W{k_1 \choose{k_i}}$,
and the reduced MEP-entropy $\frac1N \log M$ is given by,
\begin{equation}
	S_{\rm MEP} =-\sum_{i=2}^W \left[ p_i\log\left(\frac{p_i}{p_1}\right)+(p_1-p_i)\log\left(1-\frac{p_i}{p_1}\right)\right]\quad .
\label{scent}
\end{equation} 
The numerical values are, 
\begin{equation}
\begin{array}{lcl}
	S_{\rm MEP}&=&p_1\sum_{i=2}^W(1-\frac1i)\log(1-\frac1i)+p_1\sum_{i=2}^W \frac{1}{i}\log i \\
&\sim& \frac12\log W+ 1+ {\cal O}(1/\log W) \quad . 
\end{array}
\label{SX}
\end{equation}
Similarly, one can determine the probability of sampling a particular sequence $x$.
Each visit to a state $i>1$ in the sequence $x$ 
contributes to the probability of the next visit to a state $j<i$ with a factor $1/Q_{i-1}$,
whatever $j$ gets sampled. Only if $i=1$, we do not get such a renormalization factor, 
since the process restarts and all states $i$ are valid targets with probability $q_i$. It follows that
$G(k|q,N)=\prod_{i=1}^W q_i^{k_i} \prod_{j=2}^W Q_{i-1}^{-k_i}$, 
and the cross-entropy is found to be, 
\begin{equation}
	S_{\rm cross}(p|q)=-\sum_{i=1}^W p_i\log q_i + \sum_{i=2}^W p_i\log Q_{i-1}\quad.
\label{sccross}
\end{equation} 
Since terms in $S_{\rm cross}$ do not cancel terms in $S_{\rm MEP}$, we can safely identify 
$S_{\rm MEP}$ with the reduced Boltzmann entropy, $S_{\rm MEP} = s_B$.

The relative entropy of the staircase process is  
\begin{equation}
	S_{\rm rel} =S_{\rm cross}-S_{\rm MEP}\quad . 
	\label{scMEP}
\end{equation}
To get the maximum configuration we have to minimize 
$S_{\rm rel}$ with respect to $p$ under the constraint $\sum_{i=1}^W p_i=1$.
The result is derived in Appendix \ref{apB} and reads,
\begin{equation}
	p_i=p_1\frac{q_i}{Q_{i}}\quad.
\label{solfin}
\end{equation} 
For constant prior probabilities $q_i=1/W$ this yields Zipf's law
$p_i=p_1/i$, with $p_1$ a normalization constant. 

Note that the form of the MEP-entropy of SSR processes, $S_{\rm MEP}(p) =\sum_{i=2}^W s(p_1,p_i)$,
is not of trace-form, since the state $i=1$, remains entangled with every other state $j>1$. 
SSR processes violate almost all SK axioms. 
For perfectly ordered states with distributions $p_i=\delta_{ij}$ that are 
concentrated on a single state $j$, $S_{\rm MEP}(\delta_{ij})=0$. 
For the uniform distribution $p_i=1/W$, we get $S_{\rm MEP}(p)=0$.
This property has been advocated by Gell-Mann and Lloyd for functionals
measuring a so-called {\em effective complexity} \cite{GellMannLloyd,GellMannLloyd2}.
This property emerges from the fact that for SSR processes
the uniform distribution can only be obtained if the process evolves along the 
particular sequence $W\to W-1\to W-2\to\cdots\to2\to1$, which is immensely unlikely.

\subsection{Summary SSR processes}

For sufficiently large $W$ the values for entropy production $S_{\rm IT}$, of MEP-entropy 
$S_{\rm MEP}$ (reduced Boltzmann entropy $s_B$), and the 
generalized cross-entropy $S_{\rm cross}$, all yield the same numerical values,  
\begin{equation}
	S_{\rm IT} \sim S_{\rm MEP}\sim S_{\rm cross} \sim \frac12\log W+ {\cal O}(1+ 1/\log W) \quad.
\label{typval}
\end{equation}
Much of what is true for Markov processes remains true for SSR processes, which become Markovian
by restarting the process once it stops in $i=1$. 
The reduced Boltzmann entropy again measures the typical information rate of the process
and determines the amount of information that is required to optimally code typical SSR processes. 
Comparing Eq. (\ref{typval}) with entropy production of Bernoulli processes $\log W$, 
note that typical SSR processes only need half the information for encoding a message. 
It is remarkable that SSR processes, as driven dissipative systems, show enhanced compressibility.

\section{Multinomial mixture processes \label{sec5}}

\begin{table*}[ht]
  \caption{Extensive entropy, information theoretic entropy rate and maxent entropy for 
	P{\'o}lya, sample space reducing, and multinomial mixture processes. 
	$H(p)$ is defined in Eq. (\ref{shannon}) and $f(q)$ is the mixing kernel. 
	Expressions are generally valid for large $N$ and $W$.}
 	  \begin{tabular}{l | c c c} 
								& P{\'o}lya process (WTA)				& SSR process 			& multinomial mixture process 	\\
	\hline
$S_{\rm EXT}$ & $S_{1-\frac{\gamma}{1-q_j},0}$	& $S_{1,1} = H(p)$ (ensemble)	&	$S_{1,1} = H(p)$\\    
$S_{\rm IT}$  &  $ \frac{1-q_j}{\gamma}\frac1N \log N$  &  $1+ \frac12 \log W$  &  $\int_0^1 dq \, f(q) \, H(q)	$					\\  
$S_{\rm MEP}$ 	& $-\sum_i \log p_i$	& $-\sum_{i=2}^W \left[ p_i\log\left(\frac{p_i}{p_1}\right)+(p_1-p_i)\log\left(1-\frac{p_i}{p_1}\right)\right] $&	 	  
\begin{tabular}{c} 
dependes on mixing kernel \\ 
$f(q)=\mu(q)\gamma(q|\theta)\Rightarrow S_{\rm MEP}=\log(\mu(q))$
\end{tabular}
\\    
\end{tabular}
    \label{table2}
\end{table*} 

Multinomial mixture processes (MMP) can be viewed as two-step processes, 
where an urn is filled with dies with $W$-faces.  Each die may 
have individual biases $q=(q_1,\cdots,q_W)$. From this urn we draw a die,
toss it, record the outcome and put it back into the urn. 
In other words, one draws dice with biases $q$ according to some 
fixed probability density function $f(q)$ that is called the {\em mixing kernel}. 
Assume $f$ to be sufficiently smooth and non-vanishing for all states $i$.

\subsection{Entropy production and extensive entropy of multinomial mixture processes}

The MMP samples from the states $i=1,\cdots,W$ again and again. The process is stationary 
and if $f$ is smooth then $\bar W>1$.
As a consequence, the extensive entropy of such processes must be $(c,d)=(1,1)$, 
\begin{equation}
	S_{\rm EXT}(p) = S_{(1,1)}(p) = H (p) \quad ,	
\end{equation}
meaning that the extensive entropy is $H$.

MMPs are ergodic. Therefore for each set of biases $q$ in the mixture, one 
gets a typical contribution $H(q)$ to the entropy production, and 
the entropy rate of a typical sequence is given by the expectation value,
\begin{equation}
	S_{\rm IT}(x)\sim \int_0^1 dq\ f(q)\delta(1-|q|_1) H(q)\equiv \langle H\rangle_f \quad,
\end{equation}
which is nothing but the conditional entropy to draw a die with weights $q$, given that $q$ is drawn with probability $f$. 
Note that with expected frequencies are, 
\begin{equation}
	p_i=\int_0^1 dq\ f(q)\delta(1-|q|_1)q_i\equiv\langle q_i\rangle_f\quad.	
\end{equation}
It follows that in general $H(\langle q\rangle_f)>\langle H\rangle_f$,
meaning that Shannon entropy of the stationary distribution overestimates 
the information rate of the process.

\subsection{The MEP of multinomial mixture processes}

Assume a MMP with $\theta=q$. The probability to sample histogram $k$ is,
\begin{equation}
	P(k)=M(k)\int_0^1 dq f(q) \delta(|q|_1-1)\prod_{j=1}^W q_j^{k_j} \quad ,
	\label{mmproc}
\end{equation}
where $M(k)$ is the multinomial factor, $|q|_1=\sum_{i=1}^W q_i$, and $f$ is normalized, $1=\int_0^1 dq f(q) \delta(|q|_1-1)$. 
Just as in the case of the P{\'o}lya process, one might naively think that the MEP functional is $H$ plus cross-entropy terms. 
Again, this turns out to be wrong. Consider the identity, 
\begin{equation}
	\left[M(k)\frac{(N+W-1)!}{N!}\right]^{-1}=\int_0^1dq \prod_{i=1}^W q_i^{k_i}\delta(|q|_1-1)\quad.
\end{equation}
Since for a distribution $p$ with $|p|_1=1$ the function $\prod_{i=1}^W q_i^{p_i}$ is
maximal for $p=q$, we see that for large $N$,
\begin{equation}
	\prod_{i=1}^W\delta\left(q_i-\frac{k_i}{N}\right) \sim M(k)\frac{(N+W-1)!}{N!}\prod_{i=1}^W q_i^{k_i}\delta(|y|_1-1)\quad,
\label{asdelta}
\end{equation}
forms a so-called delta-sequence. 
Inserting Eq. (\ref{asdelta}) into Eq. (\ref{mmproc}) gives,
\begin{equation}
	P(k)\sim N^{1-W}f\left(\frac{k}{N}\right)\quad, 
\end{equation}
and the relative entropy of MMPs with $f(q|\theta)$ is, 
\begin{equation}
	S_{\rm rel}(p|\theta)=-\log f(p|\theta)\quad.
\end{equation}
If $S_{\rm rel}$ can be decomposed into $S_{\rm rel}=S_{\rm MEP}-\tilde S_{\rm cross}$,
depends on the mixing kernel $f$. If it factorizes $f(q|\theta)=\mu(q)\gamma(q|\theta)$, 
then $S_{\rm MEP}(q)\sim\log \mu(q)$, 
$S_{\rm cross}(q|\theta)\sim -\log\gamma(q|\theta)$ and, 
\begin{equation}
	P(k|\theta)=M(k)\int dq \delta(|q|_1-1)\prod_{j=1}^W q_j^{k_j} \frac1Ze^{S_{\rm MEP}-S_{\rm cross}} \quad , 
\end{equation}
where $Z$ is a normalization constant.

\section{Conclusions \label{sec6}}

For simple systems the concepts of thermodynamic entropy, information theoretic entropy 
and the entropy in the maximum entropy principle all lead to the same 
entropy functional $H(p)$, it is degenerate. 
The essence behind simple systems and processes rests in the fact that they are all basically 
built on multinomial Bernoulli processes. We showed that Bernoulli processes generically lead to 
$H$, whatever entropy concept is used. 
We showed in three concrete examples that this degeneracy is broken for more complex processes, 
and that the three entropy concepts lead to completely distinct functional forms. 
The entropy concepts now capture information about distinct properties of the underlying system.
The three processes studied were the P{\'o}lya process as an example for a self-reinforcing process, 
sample space reducing processes as an example of history-dependent processes with 
power law distribution functions, and multinomial mixture processes, which serve as an 
example of composed stochastic processes. The results are summarized in table \ref{table2}.  
The processes discussed here are relatively simple when compared to stochastic processes that occur  
in actual non-egodic complex adaptive systems, which often are self-reinforcing, 
path-dependent and composed of multiple dynamics. 
The main contribution of our exercise here is that it unambiguously shows that for any process that 
can not be based on, or be traced back to Bernoulli processes, one needs
to exactly specify which concept of entropy  one is talking about before it makes sense to try to compute it. 
In general the three concepts have to be computed system class by system class. 
To naively use the expression $H$ as a one-fits-all 
concept is doomed to lead to confusion and nonsense. It remains to be seen if systems 
and processes can be classified into families that share the same three faces of entropy. 

Supported by the Austrian Science Foundation FWF under the projects P29032 and I3073.

\begin{appendix}

\section{Existence of a unique extensive entropy for non-extensive systems \label{apA}}

Assume that the effective phasespace volume is given by 
\begin{equation}
	\hat W(N)\equiv \prod_{n=1}^N \bar W(X_n) \quad .
\end{equation}
Since $\hat W(N)$ is monotonically increasing in $N$ an inverse function $L_X$ exists 
such that $L_X(\hat W(N))=N$, and a unique extensive trace-form functional can be found, 
\begin{equation}
	S_{\rm EXT}(p)=\sum_{x\in\Omega^N}s(p(x)) \quad . 
	\label{traceform}
\end{equation}
Here $q(x)$ is the probability to sample path $x$, such that for sequences $x(N)$ one obtains 
\begin{equation}
	S_{\rm EXT}(q(x(N)))=Ns_0\quad , 
	\label{extensive}
\end{equation}
with $s_0=\tilde W(1)s\left(1/\tilde W(1)\right)$.
If we look at a reference process, where the path probabilities 
$q(x)$ are uniformly concentrated on $\hat W(N)$ paths, it follows that,
\begin{equation}
	\sum_{x\in\Omega^N}s(q(x))\sim \hat W(N)s\left(\frac1{\hat W(N)}\right)\quad .
\label{extensive2}
\end{equation}
Clearly, $\hat W(N)s\left(\frac1{\hat W(N)}\right)=Ns_0$ is exactly solved by, 
\begin{equation}
	s(x)=s_0 \, x \, L_X\left(\frac1x\right)\quad.
	\label{extent}
\end{equation}

\section{Solving the MEP for SSR processes \label{apB}}

To maximize the MEP of the staircase process, Eq. (\ref{scMEP}), with respect to the probabilities 
$p$ under the constraint $\sum_{i=1}^W p_i=1$, where $W$ is the number of the possible states, we may proceed as follows. 
The staircase MEP requires to solve $\delta\left(\psi(p|q,N)-\alpha(\sum_{i=1}^W p_i-1)\right)=0$, 
where $\psi =S_{\rm MEP}-S_{\rm cross}= -S_{\rm rel}$ of the SSR process and 
$\alpha$ is the Lagrange multiplier guaranteeing the constraint. This means that every derivative 
of the constrained functional with respect to $p_i$, must be zero. 
For $i>1$ one gets,  
\begin{equation}
	p_i=\frac{p_1}{1+\zeta\frac{Q_{i-1}}{q_i}}\quad,
\label{appA:soligt1}
\end{equation} 
where $\zeta=\exp(\alpha)$.
Similarly, for $i=1$ one finds,
\begin{equation}
	q_1=\zeta\exp\left(\sum_{i=2}^W\log\left(1-\frac{p_i}{p_1}\right)\right)\quad.
\label{appA:solieq1}
\end{equation} 
Solving these two equations self-consistently one finds (at least numerically) that $\zeta=1$,
and the solution of the MEP is, 
\begin{equation}
	p_i=p_1\frac{q_i}{Q_{i}}\quad . 
\label{appA:solfin}
\end{equation}

\end{appendix}

\end{document}